\documentclass[a4paper]{jpconf}
\usepackage{graphicx}
\begin{document}
\title{Physics reach of the ESSnuSB experiment}

\author{Monojit Ghosh\footnote{On behalf of the ESSnuSB collaboration}}

\address{Center of Excellence for Advanced Materials and Sensing Devices, Ruder Bo\v{s}kovi\'c Institute, 10000 Zagreb, Croatia} 
\address{School of Physics, University of Hyderabad, Hyderabad - 500046, India}

\ead{monojit$\_$rfp@uohyd.ac.in}

\begin{abstract}
ESSnuSB is a unique future proposed long-baseline experiment in Sweden to study neutrino oscillation by probing the second oscillation maximum. In this proceeding, we update the flux and efficiencies and re-calculate the sensitivity of ESSnuSB in the standard three flavour scenario. We find that it has excellent sensitivity to the Dirac CP phase $\delta_{\rm CP}$, moderate sensitivity to the mass hierarchy of the neutrinos and limited sensitivity to measure the octant of the atmospheric mixing angle $\theta_{23}$. We also find that it has a very good sensitivity to constrain the atmospheric mass squared difference $|\Delta m^2_{31}|$.
\end{abstract}

\section{Introduction}

Among the six parameters ($\theta_{12}$, $\theta_{13}$, $\theta_{23}$, $\Delta m^2_{21}$, $\Delta m^2_{31}$ and $\delta_{\rm CP}$) that describe the neutrino oscillation in standard three flavour scenario, the current unknowns are: (i) neutrino mass hierarchy which can be either normal i.e.,$\Delta m^2_{31} < 0$ or inverted i.e., $\Delta m^2_{31} < 0$, (ii) the octant of $\theta_{23}$ which can be either higher i.e., $\theta_{23} > 45^\circ$ or lower i.e., $\theta_{23} < 45^\circ$ and (iii) $\delta_{\rm CP}$ \cite{Esteban:2020cvm}. ESSnuSB \cite{ESSnuSB:2013dql,Wildner:2015yaa,Blennow:2019bvl} is one of the proposed long-baseline experiments in Sweden which aims to measure the CP phase $\delta_{\rm CP}$ by probing the second oscillation maximum. As the variation of the oscillation probability with respect to $\delta_{\rm CP}$ is higher in the second oscillation maximum as compared to the first oscillation maximum, this experiment can measure $\delta_{\rm CP}$ with very good precision. 

In this proceedings, we have updated the calculation of flux and efficiencies and re-calculated the sensitivity of ESSnuSB to measure the unknowns mentioned above in the standard three flavour scenario. This proceedings is organized as follows.  In the next section we will give the details of the simulation which we used in our calculation. Then we will present our results. Finally we will summarize and conclude. 

\section{Simulation Details}

We consider a water Cherenkov detector of fiducial volume 538 kt located either at a distance of 540 km or 360 km from the neutrino source. Neutrino beam is produced by a powerful linear accelerator (linac) capable of delivering $2.7 \times 10^{23}$ protons on target per year having a beam power of 5 MW with proton kinetic energy of 2.5 GeV. The fluxes and the event selection for the Far Detectors are estimated using full Monte Carlo simulations specific to the ESSnuSB conditions. These fluxes and detector response with efficiencies are then incorporated in GLoBES \cite{Huber:2004ka,Huber:2007ji} to calculate event rates and $\chi^2$. We have considered the systematic errors on the overall normalization of the expected number of detected events at the Far Detectors: $5\%$ for signal and $10\%$ for background. No systematic effects on the shape of the detected energy spectrum have been implemented. The systematic errors are considered to be the same for appearance and disappearance channels for both neutrinos and antineutrinos. We have considered a total run-time of 10 years which is divided into 5 years of neutrino beam and 5 years of antineutrino beam. The configurations are same for both baseline options of ESSnuSB. 

\section{Results}

\begin{figure*}[t]
\begin{center}
\includegraphics[width=0.57\textwidth]{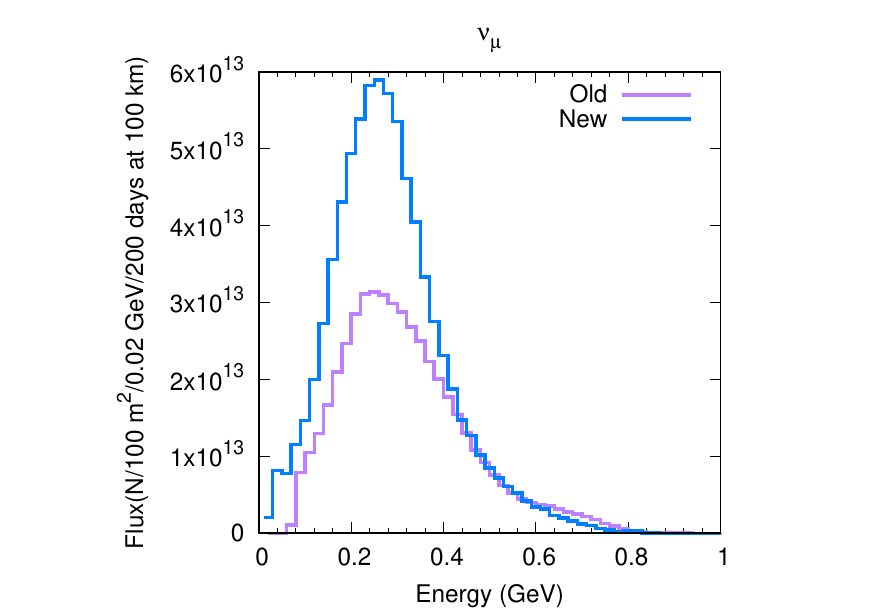} 
\hspace{-1.0in}
\includegraphics[width=0.57\textwidth]{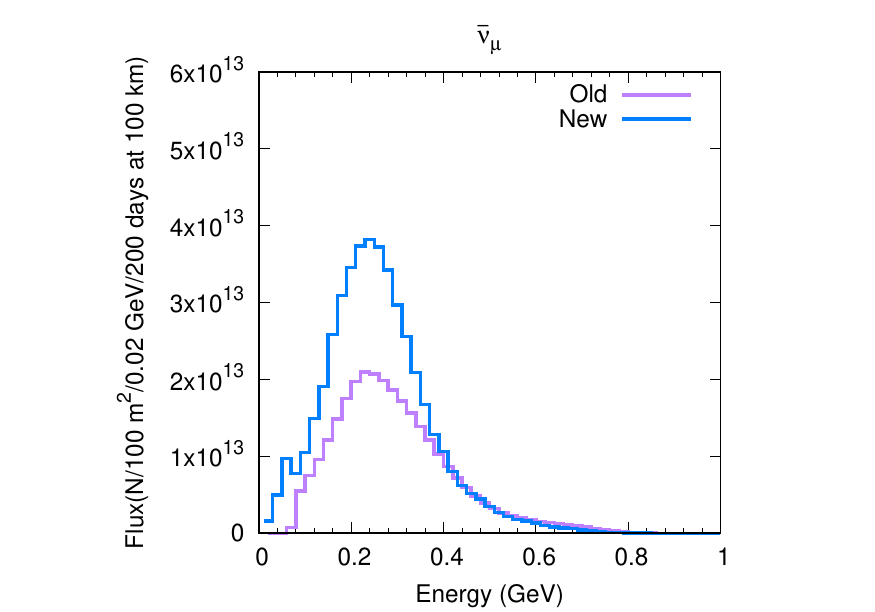} \\
\end{center}
\caption{Neutrino fluxes as a function of energy. The left panel is for positive polarity and the right panel is for negative polarity.}
\label{fig_flux}
\end{figure*}

\begin{figure*}[t]
\begin{center}
\includegraphics[width=0.57\textwidth]{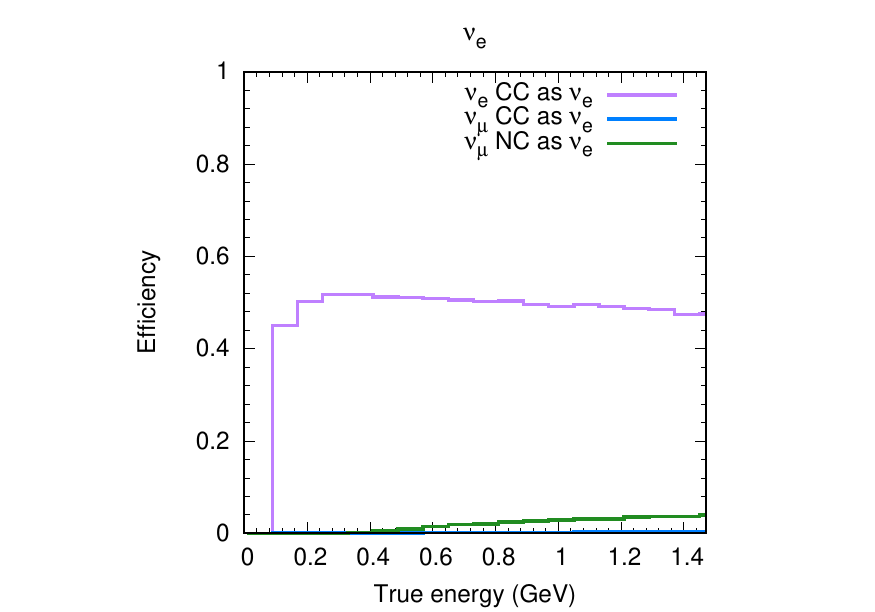} 
\hspace{-1.0in}
\includegraphics[width=0.57\textwidth]{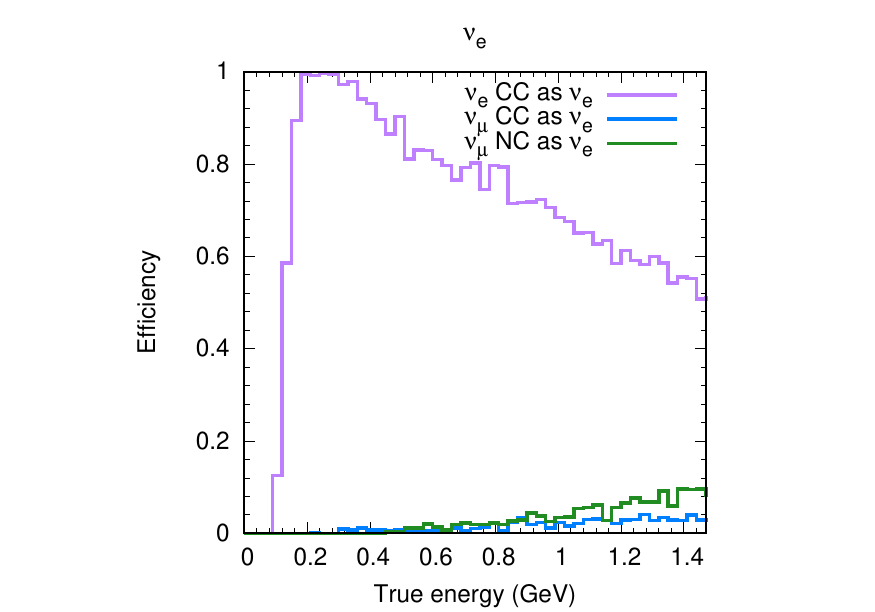} \\
\includegraphics[width=0.57\textwidth]{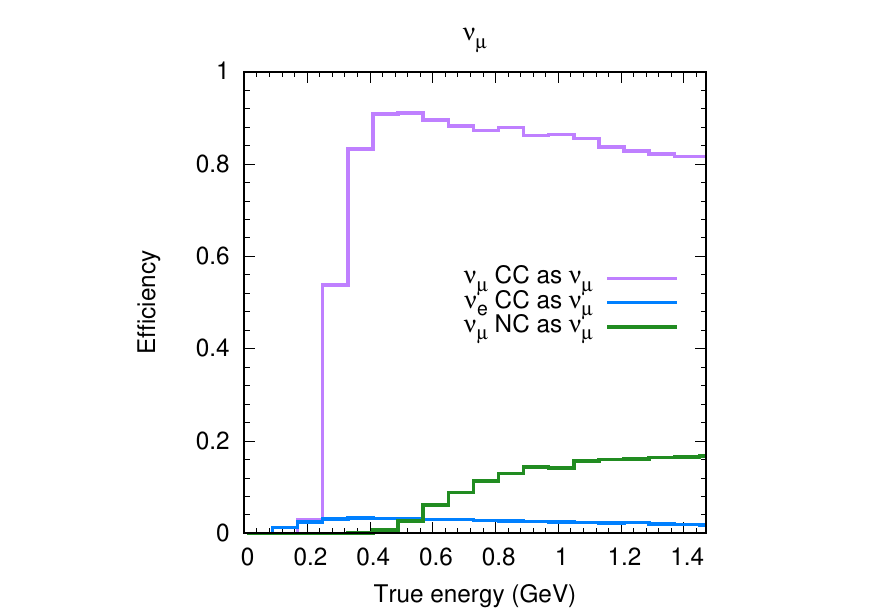} 
\hspace{-1.0in}
\includegraphics[width=0.57\textwidth]{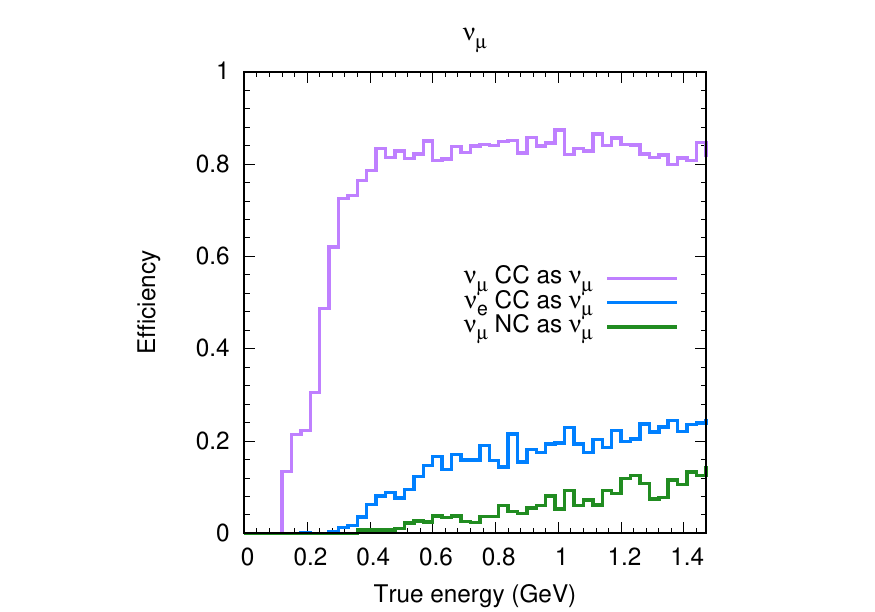} \\
\end{center}
\caption{Efficiencies as a function of energy. The left column is for previous selection and the right column is for updated selection.}
\label{fig_eff}
\end{figure*}

\begin{figure*}
\begin{center}
\includegraphics[width=0.4\textwidth]{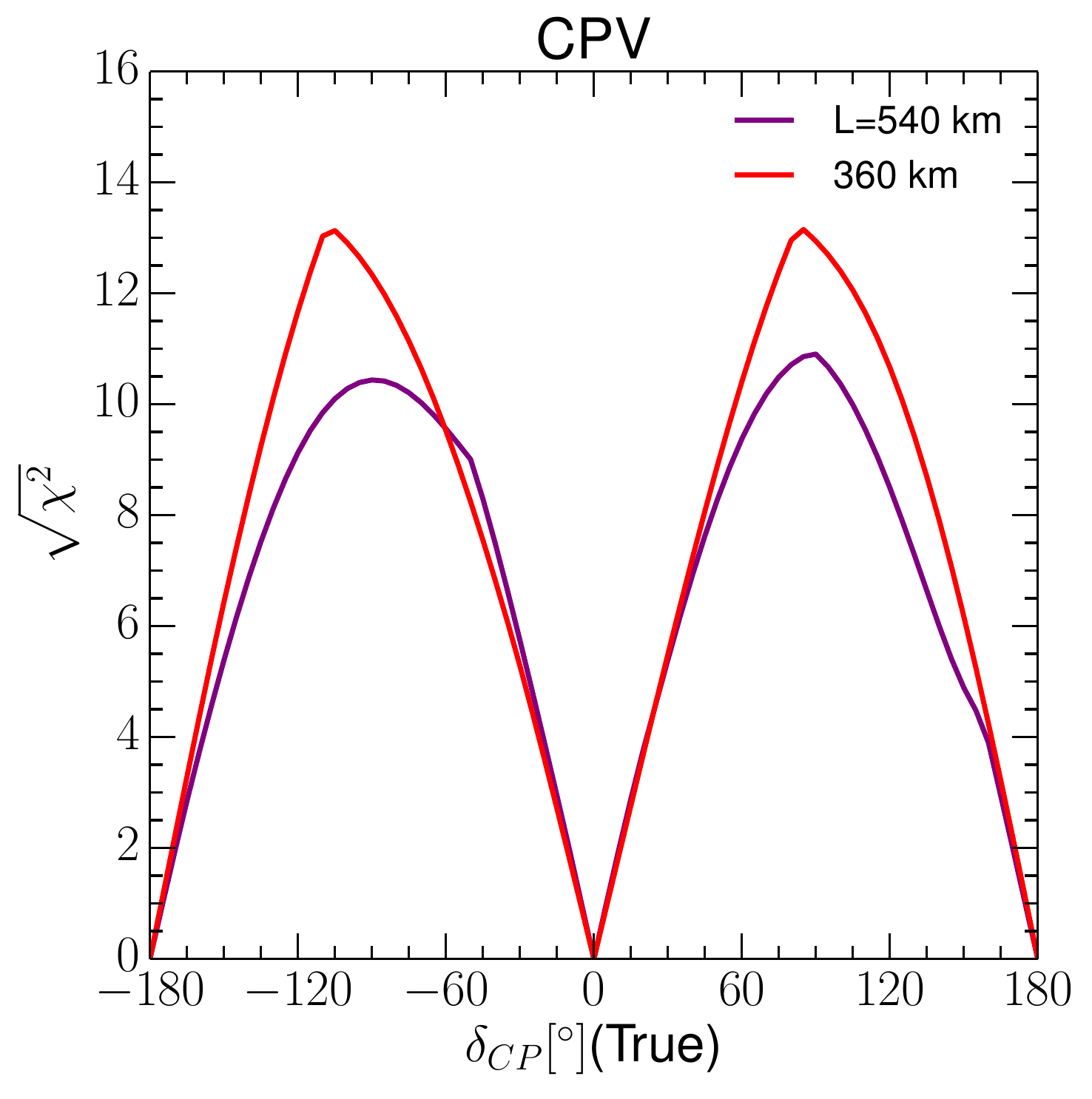} 
\includegraphics[width=0.4\textwidth]{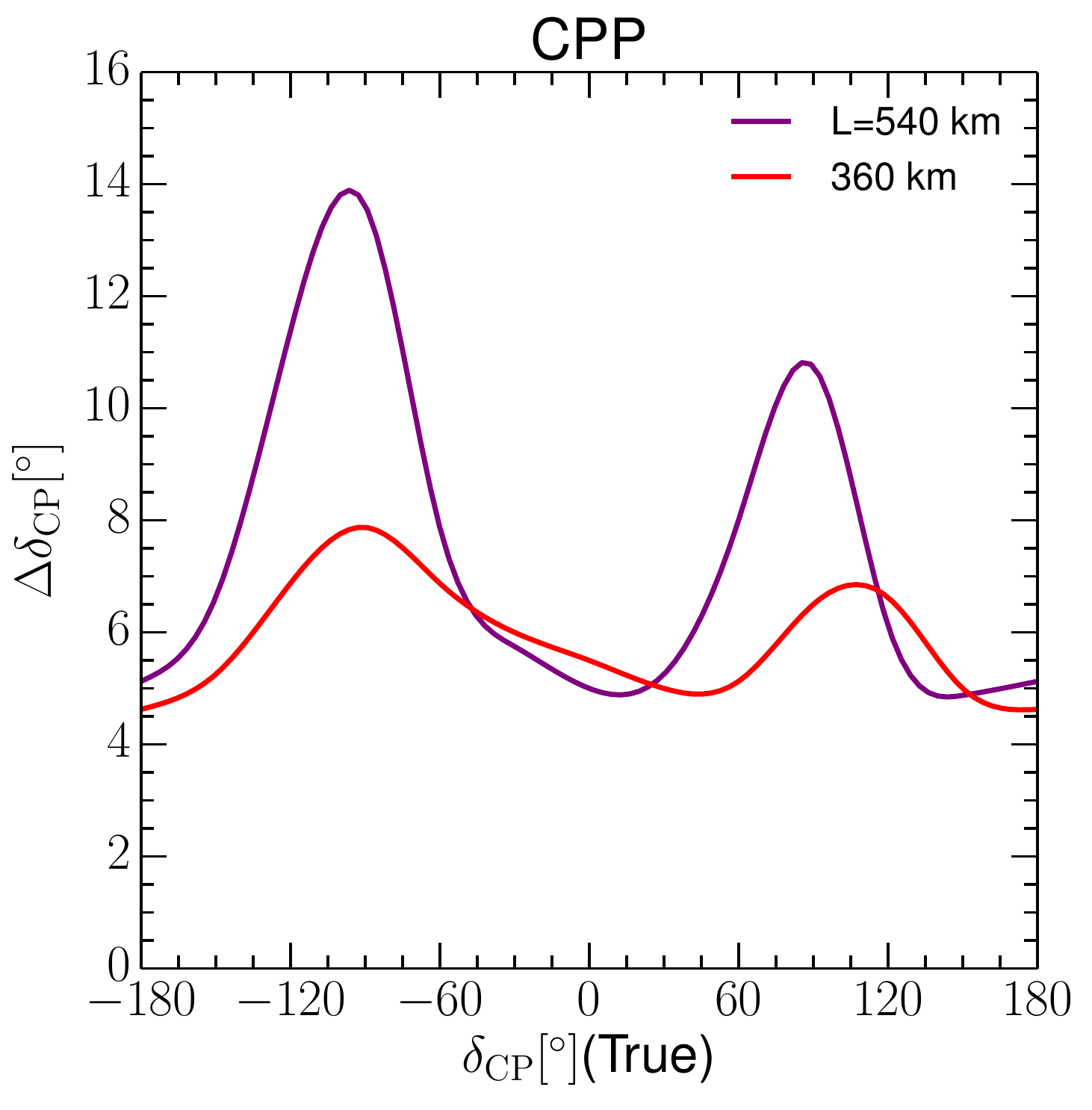} \\
\end{center}
\caption{CP violation sensitivity (left panel) and CP precision sensitivity (right panel) as a function of $\delta_{\rm CP}$ true.}
\label{fig_cp}
\end{figure*}

\begin{figure*}
\begin{center}
\includegraphics[width=0.31\textwidth]{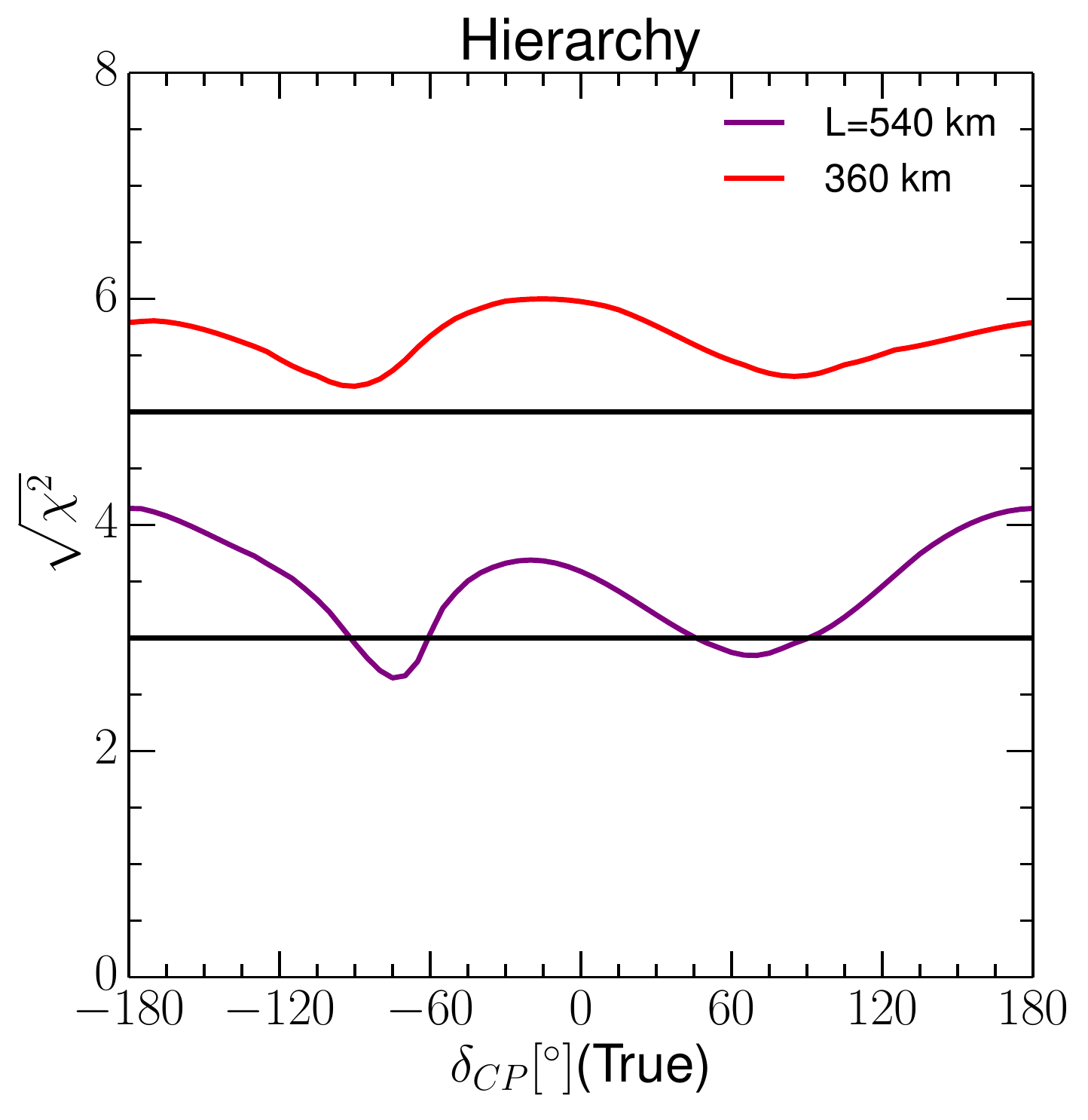} 
\includegraphics[width=0.33\textwidth]{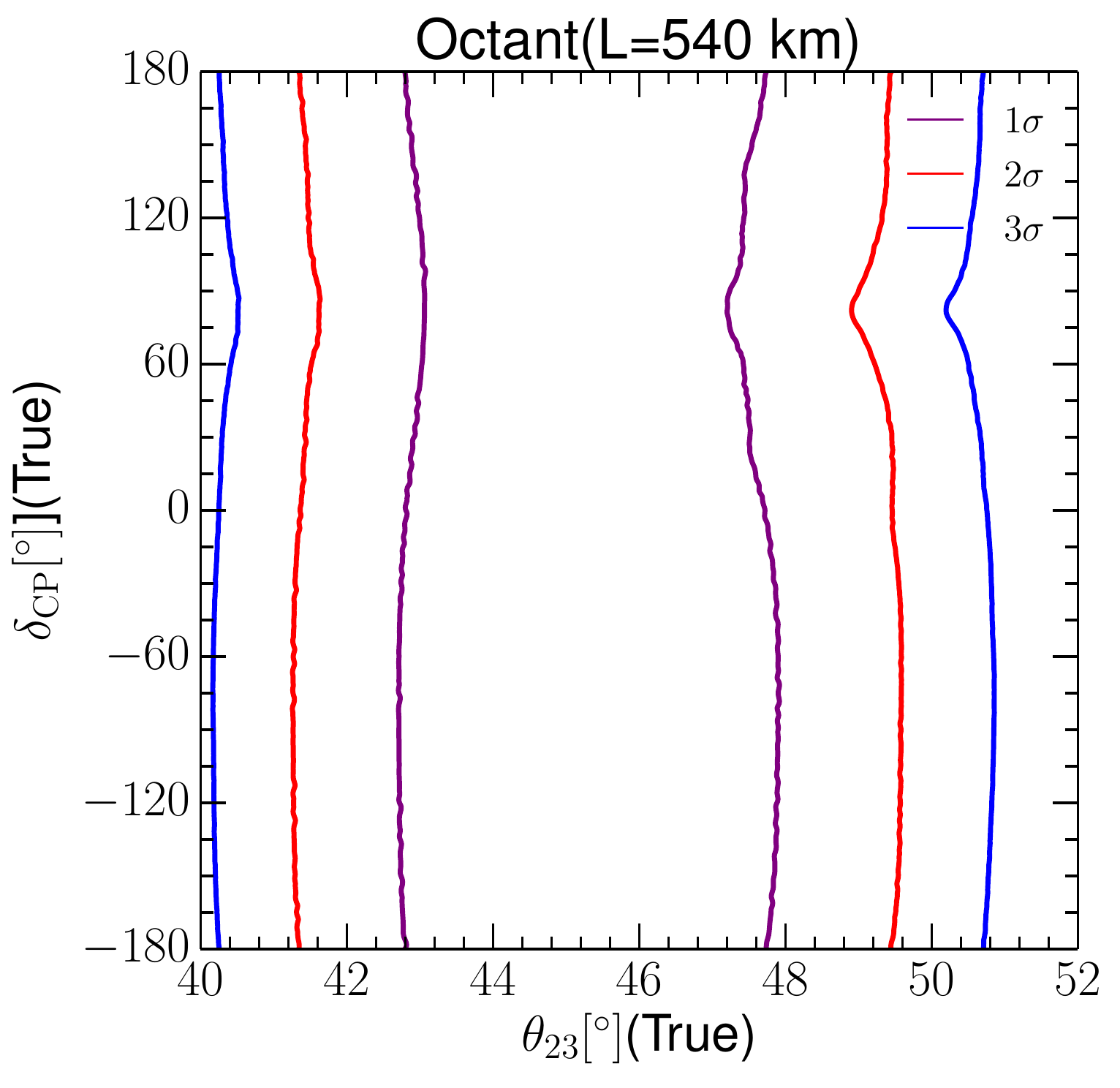}
\includegraphics[width=0.33\textwidth]{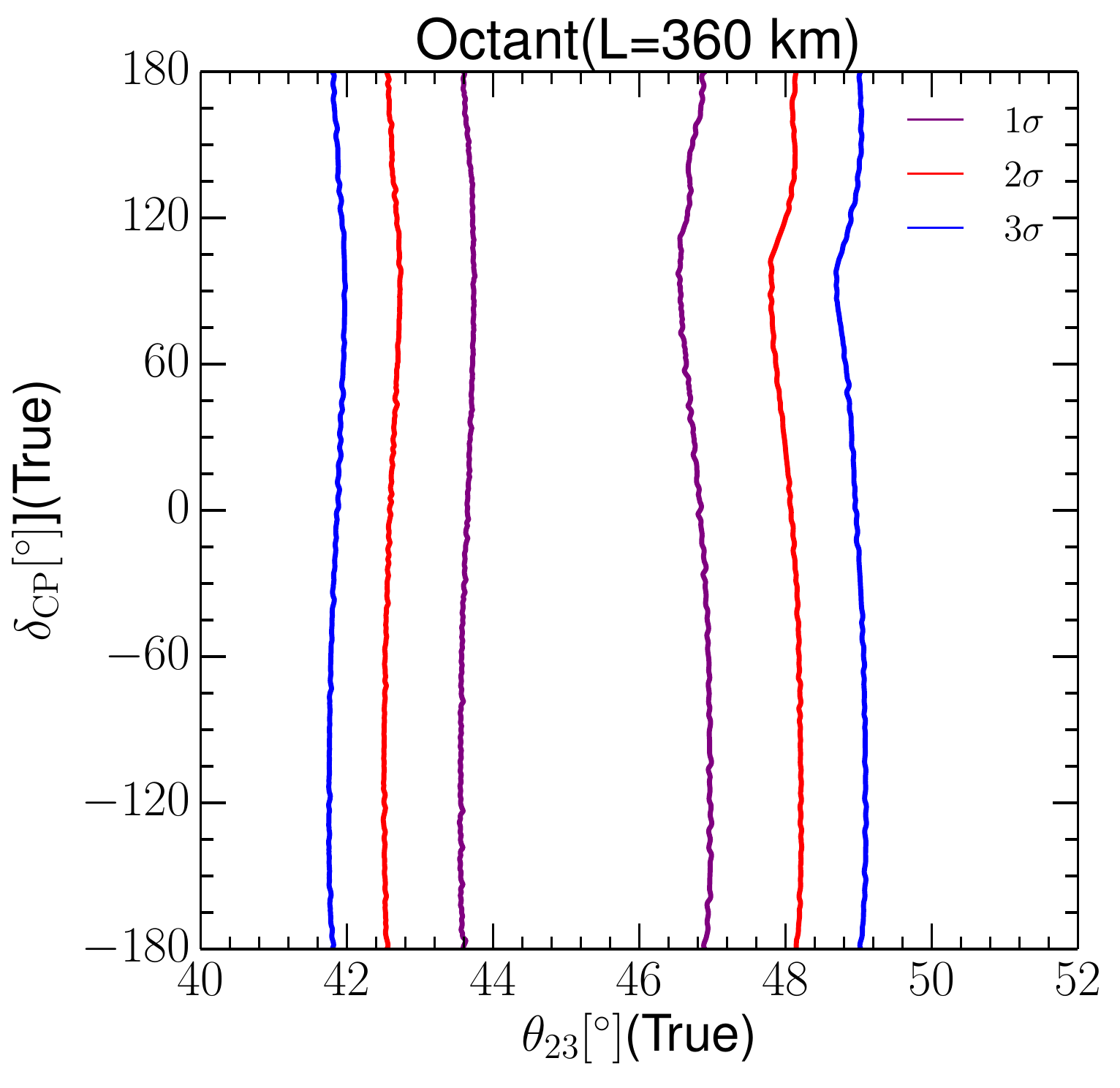}
\end{center}
\caption{Hierarchy sensitivity (left panel) as a function of $\delta_{\rm CP}$ true and octant sensitivity (middle and left panel) in the $\theta_{23}$ true - $\delta_{CP}$ true plane.}
\label{fig_hier_oct}
\end{figure*}

\begin{figure*}
\begin{center}
\includegraphics[width=0.4\textwidth]{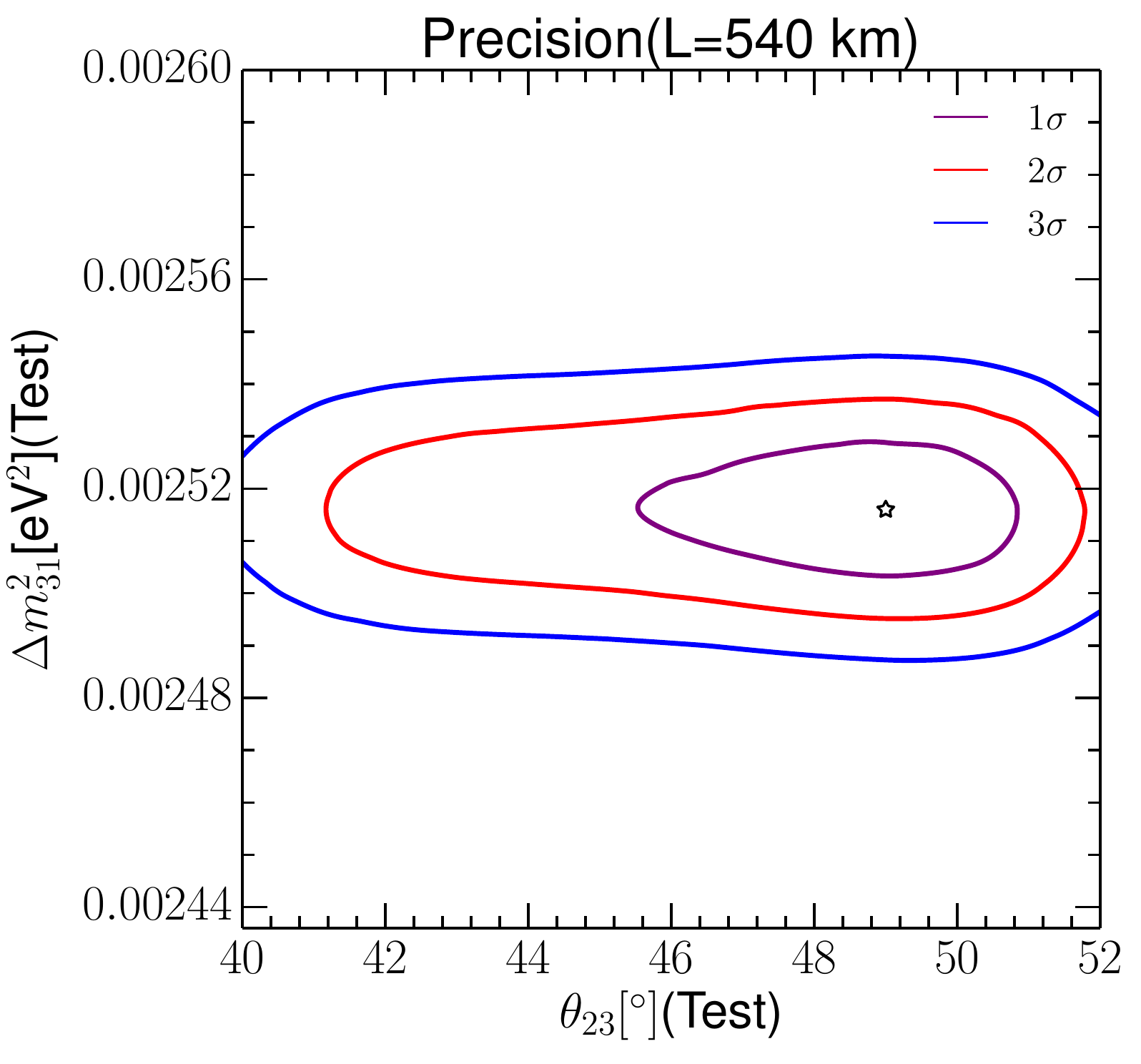} 
\includegraphics[width=0.4\textwidth]{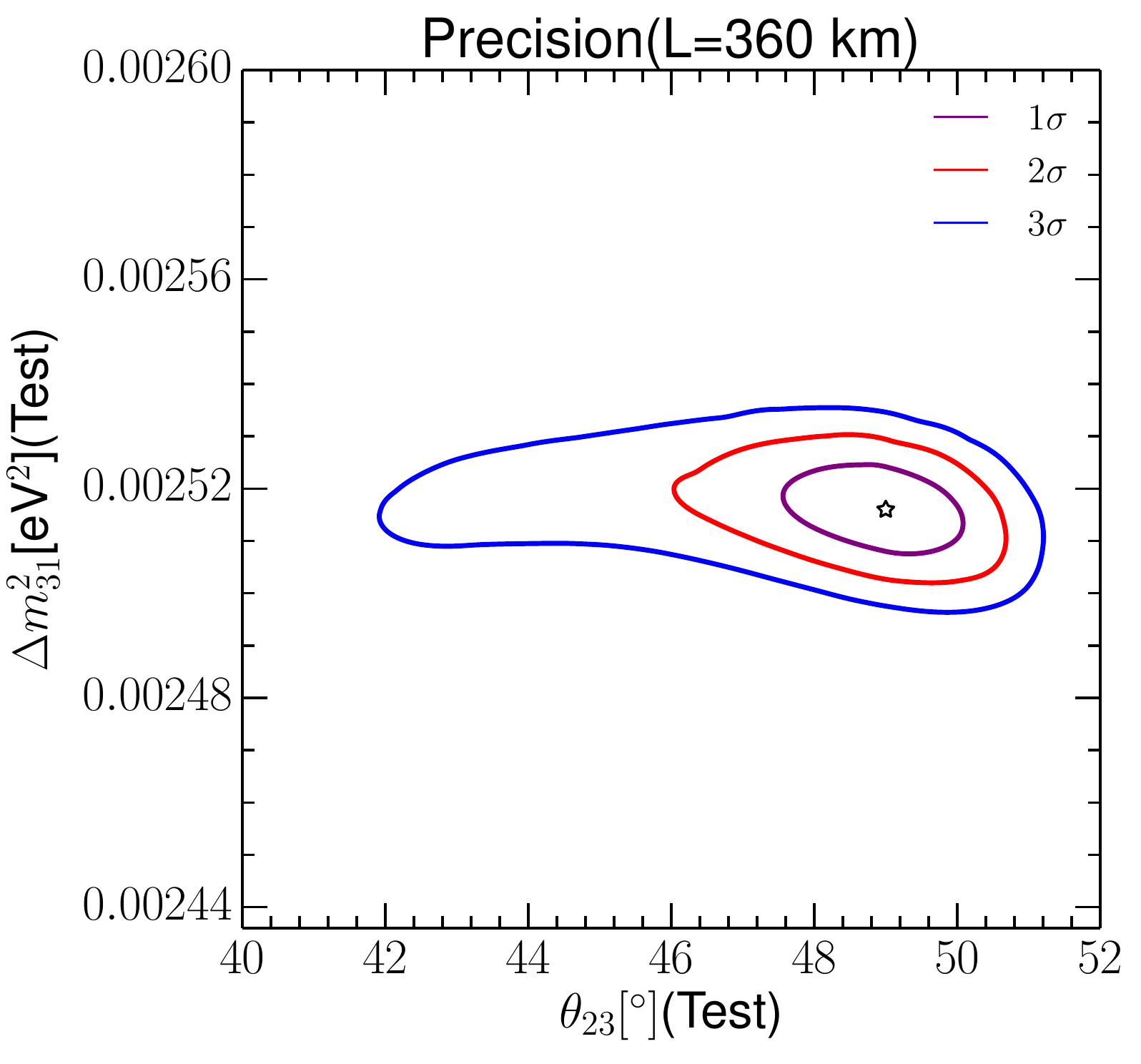} \\
\end{center}
\caption{Precision of the atmospheric mixing parameters. Left panel is for 540 km and the right panel is for 360 km.}
\label{fig_prec}
\end{figure*}

In Fig. \ref{fig_flux}, we have presented the muon flux as a function of energy. The left panel is for positive polarity and the right panel is for negative polarity. In each panel the purple curve corresponds to the old flux as given in \cite{MEMPHYS:2012bzz} and the blue curve corresponds to the updated flux used in this analysis. From this panel we understand that there is a significant improvement in the updated flux. 

In Fig. \ref{fig_eff}, we have plotted the efficiency as a function true energy. The left column is for the previous selection as given in \cite{MEMPHYS:2012bzz} and the right column is for the updated selection which is used in the present calculation. From these panels we understand that the updated $\nu_e$ signal efficiency (purple curve in the top row) is much higher as compared to the previous selection, whereas the updated $\nu_\mu$ efficiency (purple curve in the bottom row) is somewhat similar to the previous selection. As CP sensitivity mainly comes from the appearance channel, we expect a significant improvement in the CP sensitivity with the current selection.  

In Fig. \ref{fig_cp}, we have plotted the CP sensitivity of ESSnuSB with the updated flux and updated efficiencies. The left panel is for CP violation and the right panel is for CP precision. In each panel, the purple curve corresponds to 540 km and the red curve corresponds to 360 km. For CP violation, we understand that for $\delta_{CP} = \pm 90^\circ$, we have around $14 \sigma$ sensitivity for 360 km baseline and around $10 \sigma$ sensitivity for 540 km baseline. For CP precision, we understand that the $1 \sigma$ precision of $\delta_{\rm CP}$  is around $5^\circ$ if the true values of $\delta_{\rm CP}$ are around $0^\circ$ or $180^\circ$ for both baseline options. However, for $\delta_{\rm CP} = -90^\circ$, the error is around $14^\circ$ for the baseline option of 540 km and only $7^\circ$ for the baseline option of 360 km. CP sensitivity for 360 km is higher because of the larger statistics at the smaller baseline.

In Fig. \ref{fig_hier_oct}, we present the hierarchy and octant sensitivity of ESSnuSB. In the left panel we present the hierarchy sensitivity as a function of $\delta_{\rm CP}$ (true). The purple curve corresponds to the baseline option of 540 km and the red curve corresponds to the baseline option of 360 km. The black horizontal lines correspond to the benchmark of $3\sigma$ and $5\sigma$ sensitivity, respectively. From this panel we understand that for the baseline option of 540 km, one can have a $3 \sigma$ hierarchy sensitivity except for $\delta_{\rm CP} = \pm 90^\circ$, and for the baseline option of 360 km one can have a hierarchy sensitivity of $5 \sigma$ for all the values of $\delta_{\rm CP}$. The hierarchy sensitivity is higher for 360 km is because of higher matter effect. In the middle and left panels we present the octant sensitivity in the $\theta_{23}$ (true) vs $\delta_{\rm CP}$ (true) plane. The middle panel is for the baseline option of 540 km and the right panel is for the baseline option of 360 km. In each panel the purple/red/blue curve corresponds to the 1$\sigma$/2$\sigma$/3$\sigma$ contours, respectively. In these panels, the region around $\theta_{23}=45^\circ$ shows the values of $\theta_{23}$ for which the octant cannot be determined at that given C.L. From these panels we see that the octant sensitivity of ESSnuSB is limited. In these panels the sensitivity of 360 km is slightly better than the 540 km. 

Finally, in Fig. \ref{fig_prec}, we plot the precision measurement of the atmospheric mixing parameters of ESSnuSB in the $\theta_{23}$ (test) - $\Delta m^2_{31}$ (test) plane. The left panel is for the baseline option of 540 km and the right panel is for the baseline option of 360 km. In each panel, the purple/red/blue curve corresponds to the 1$\sigma$/2$\sigma$/3$\sigma$ C.L. contours, respectively. The measured central values of $\theta_{23}$ and $\Delta m^2_{31}$ are indicated by a star. From these panels we understand that the capability of ESSnuSB to constrain $\Delta m^2_{31}$ is quite good, while its capability to constrain $\theta_{23}$ is limited. In terms of the precision of the atmospheric mixing parameters, the capability of the 360 km baseline is significantly better than the 540 km baseline. 

We have generated all the figures for normal mass hierarchy of the neutrinos. The true value of the oscillation parameters are: $\theta_{12} =33.44^\circ$ , $\theta_{13} = 8.57^\circ$, $\theta_{23} = 49.2^\circ$, $\Delta m^2_{21} = 7.42 \times 10^{-5}$ eV$^2$, $\Delta m^2_{31} = 2.517 \times 10^{-3}$ eV$2$ and $\delta_{\rm CP} = -163^\circ$. We have minimzed the parameters $\theta_{23}$ and $\delta_{\rm CP}$ in the test spectrum of the $\chi^2$. 

\section{Summary and Conclusion}

In this proceeding, we have studied the physics reach of the ESSnuSB experiment in the standard three flavour scenario, with the updated flux and updated selection. We have shown that the current updated flux are significantly better than the previous flux. Whereas the current $\nu_e$ event selection is much better than the previous $\nu_e$ selection but the current $\nu_\mu$ selection is comparable with the previous selection. In our analysis we find that, CP violation discovery sensitivity is $10\sigma\ (13\sigma)$ for the baseline option of 540 km (360 km) at $\delta_{\rm CP} = \pm 90^\circ$.  Regarding CP precision, the $1\sigma$ error associated with $\delta_{\rm CP} = 0^\circ$ is around $5^\circ$ for both of the baseline options and the error associated with $\delta_{\rm CP} = -90^\circ$ is around $14^\circ$ $(7^\circ)$ for the baseline option of 540 km (360 km). For neutrino mass hierarchy, one can achieve $3\sigma$ sensitivity for the 540 km baseline except for the true values of $\delta_{\rm CP} = \pm 90^\circ$ and $5\sigma$ sensitivity for the 360 km baseline for all values of $\delta_{\rm CP}$. The values of $\theta_{23}$ for which the octant can be determined at $3 \sigma$ is $\theta_{23} > 51^\circ$ ($\theta_{23} < 42^\circ$ and $\theta_{23} > 49^\circ$) for the baseline of 540 km (360 km). Regarding the precision of the atmospheric mixing parameters, the allowed values at $3 \sigma$ are: $40^\circ < \theta_{23} < 52^\circ$ ($42^\circ < \theta_{23} < 51.5^\circ$) and $2.485 \times 10^{-3}$ eV$^2 < \Delta m^2_{31} < 2.545 \times 10^{-3}$ eV$^2$ ($2.49 \times 10^{-3}$ eV$^2 < \Delta m^2_{31} < 2.54 \times 10^{-3}$ eV$^2$) for the baseline of 540 km (360 km).  Among the two baseline options, 360 km provides the better sensitivity.
For more details see Ref. \cite{ESSnuSB:2021lre} on which this proceeding is based upon. 

\section*{Acknowledements}

This project received funding from the European Union’s Horizon 2020 research and innovation programme under grant agreement No. 777419. This is also partly funded by Ramanujan Fellowship of SERB, Govt. of India, through grant no: RJF/2020/000082.

\end{document}